\documentclass{iopart}
\usepackage{iopams}
\newcommand{\eqref}[1]{(\ref{#1})}

\usepackage{eucal}

\newtheorem{theorem}{Theorem}[section]

\def\al{\alpha}
\def\ga{\gamma}
\def\be{\beta}
\def\ep{\epsilon}
\def\om{\omega}
\def\la{\lambda}
\def\bom{\boldsymbol{\omega}}
\def\bOm{\boldsymbol{\Omega}}

\def\bG{\boldsymbol{\Gamma}}

\def\bomu^#1_#2{\bom^{#1}_{\ #2}}
\def\bOmu^#1_#2{\bOm^{#1}_{\ #2}}
\def\Gu^#1_#2{\Gamma^{#1}_{\ #2}}
\def\tGu^#1_#2{\tilde{\Gamma}^{#1}_{\ #2}}
\def\R^#1_#2{R^{#1}_{\ #2}}
\def\tR^#1_#2{\tilde{R}^{#1}_{\ #2}}
\def\M^#1_#2{M^{#1}_{\ #2}}
\def\C^#1_#2{C^{#1}_{\ #2}}
\def\X^#1_#2{X^{#1}_{\ #2}}
\def\u^#1_#2{u^{#1}_{\ #2}}
\def\vu^#1_#2{v^{#1}_{\ #2}}
\def\Up^#1_#2{\Upsilon^{#1}_{\ #2}}
\def\tUp^#1_#2{\tilde{\Upsilon}^{#1}_{\ #2}}

\def\sec{2${}^{\mathrm{nd}}$}

\newcommand{\Rset}{\mathbb{R}}

\newcommand{\so}{\mathfrak{so}}

\newcommand{\fg}{\mathfrak{g}}

\newcommand{\CH}{\mathrm{CH}}

\newcommand{\mM}{\mathsf{M}}

\newcommand{\tGamma}{\tilde{\Gamma}}

\newcommand{\cF}{\mathcal{F}}

\newcommand{\cL}{\mathcal{L}}

\newcommand{\ve}{\mathbf{e}}

\newcommand{\vm}{{\boldsymbol{m}}}
\newcommand{\bvm}{{\bar{\vm}}}
\newcommand{\vell}{\boldsymbol{\ell}}
\newcommand{\vn}{\boldsymbol{n}}

\begin{document}

\title{On curvature homogeneous 4D Lorentzian manifolds}
\author{R. Milson, N. Pelavas}
\address{Dept. Mathematics and Statistics, Dalhousie University\\
  Halifax NS B3H 3J5, Canada}
\eads{\mailto{rmilson@dal.ca}, \mailto{pelavas@mathstat.dal.ca}}

\begin{abstract}
  We prove that a four-dimensional Lorentzian manifold that is
  curvature homogeneous of order $3$, or $\CH_3$ for short, is
  necessarily locally homogeneous. We also exhibit and classify
  four-dimensional Lorentzian, $\CH_2$ manifolds that are not
  homogeneous.  
\end{abstract}

\section{Introduction}
A pseudo-Riemannian manifold $(M,g_{ij})$ is locally homogeneous if
the Lie algebra of Killing vector fields spans the tangent space at
all points of $M$.  We say that $M$ is curvature homogeneous of order
$s$, or equivalently that $M$ belongs to class $\CH_s$ if the
components of the curvature tensor and its first $s$ covariant
derivatives are constant relative to some local frame.

Local homogeneity implies curvature homogeneity of all orders.
Remarkably, the converse is also true, in some fashion.  Let us say
that a manifold is \emph{properly} $\CH_s$ if it belongs to class
$\CH_s$, but is not locally homogeneous. Singer showed that there does
not exist a proper $\CH_s$ manifold if $s$ is larger than a certain
invariant, called the Singer index \cite{PoSp,singer}.  More
precisely, we have the following.
\begin{theorem}[Singer's criterion]
  \label{thm:SC}
  Let $(M,g_{ij})$ be a $\CH_s$ manifold of signature $(p,q)$.  Let
  $\fg_r\subseteq\fg_{-1}:=\so(p,q),\; r=0,1,\ldots, s$ be the
  subalgebra that leaves invariant the arrays of constants
  $\R^a_{b_1b_2 b_3}, \nabla_c \R^a_{b_1 b_2 b_3}, \ldots,
  \nabla_{c_1\ldots c_r}\R^a_{b_1 b_2 b_3}$. If $\fg_{s-1} = \fg_{s}$,
  then $M$ is locally homogeneous.
\end{theorem}

Many examples of proper $\CH_0$ Riemannian \cite{Vanhecke92} and
Lorentzian manifolds \cite{Bueken97} are known.  In the case of
three-dimensional curvature homogeneous geometries, classification
results are available \cite{BuDj00,KoVl98}.  See \cite{GiNi05} for
additional references, and for examples of higher-dimensional proper
$\CH$ manifolds of general signature.  In 4 dimensions, there are no
proper $\CH_1$ Riemannian manifolds\cite{SeSuVa92}, but this is not
true in the Lorentzian\cite{BuVa97} and the neutral-signature
setting\cite{DuGiNi04}.  At this point, no examples of
four-dimensional, proper $\CH_2$ geometries seem to be
known\cite{gilkey04}, leaving open the question of the value of the
Singer index for four-dimensional, Lorentzian, curvature-homogeneous
manifolds.

In this note we show that there are no four-dimensional, proper
$\CH_3$ Lorentzian manifolds; see section 3.  We also exhibit and
classify all four-dimensional, proper $\CH_2$ Lorentzian manifolds;
see eqns. \eqref{eq:ch2om1}-\eqref{eq:ch2om4}.  

Our approach is to formulate the field equations for a $\CH$ geometry
as an exterior differential system (EDS) on the second-order frame
bundle\cite{EsWa89}.  The integrability constraints manifest as
algebraic constraints (torsion in EDS parlance) on the curvature
scalars.  The classification is accomplished by deriving torsion-free
configurations for the $\CH$ field equations corresponding to various
algebraic types of the curvature tensor, and by using Singer's
criterion to rule out the homogeneous subcases.  See \cite{IvLa03} for
an introduction to EDS theory, see \cite{Estabrook87,Kobayashi72} for
a discussion of higher-order frames.  See \cite{Estabrook06,DoGe04}
for some recent applications of EDS theory to relativity.  We will use
this method to classify proper $\CH_1$ four-dimensional, Lorentz
geometries in a forthcoming publication.

\section{The EDS for curvature-homogeneous geometries}

Let $M$ be an $n$-dimensional, analytic manifold, and let $g_{ab}$ be
a fixed, non-degenerate bilinear form.  Henceforth, we use $g_{ab}$ to
lower frame indices, which we denote by $a,b,c=1,\ldots, n$. We use
$u^i$ to denote local coordinates, where $i,j=1,\ldots, n$ are
coordinate indices.  Even though we are interested in the case of
$n=4$ and $g_{ab}$ of Lorentzian signature, much of the underlying
theory can be given without these assumptions.

Pseudo-Riemannian structure can be specified in two ways: as a metric,
or as a $g_{ab}$-orthogonal moving frame.  In the second approach,
instead of using $g_{ij}$ and $\Gu^i_{jk}$ as unknowns in the field
equations, the field variables are scalars
\[ \u^i_a,\; \det(\u^i_a) \neq 0,\quad \u^i_{bc} = \u^i_{cb},\] the
$\frac{1}{2}n^2(n+3)$ components of a \sec order moving frame
\cite{Kobayashi72}.  The metric and connection can then be recovered as
\begin{eqnarray*}
  g_{ij} &=&   \vu^a_i \vu^b_j \,g_{ab},  \\
  \Gu^i_{jk} &=& -\vu^b_j \vu^c_k u^i{}_{bc},
\end{eqnarray*}
where $\vu^a_i \u^i_b = \delta^a_{\ b}$ is the inverse matrix of
coframe components.  It will be convenient to formulate the field
equations in terms of $1$-form  variables
\begin{eqnarray}
  \label{eq:omi}
  \bom^a &=& \vu^a_i\, du^i,\\
  \label{eq:omij}
  \bomu^a_b &=&\vu^a_i (d \u^i_b - \u^i_{bc}\, \bom^c),
\end{eqnarray} 
which represent the coframe and connection
1-forms.
The equations
\begin{eqnarray}
  \label{eq:taut1}
  &d\bom^a + \bomu^a_b\wedge \bom^b=0,\\
  \label{eq:taut2}
  &(d\bomu^a_b + \bomu^a_c\wedge \bomu^c_b) \wedge \bom^b = 0,
\end{eqnarray}
the 1st structure equations and algebraic Bianchi relations, follow as
tautologies.

Let $G= \{ \X^a_b : X_{ca} \X^c_b = g_{ab}\}$ and $\fg = \{ \M^a_b 
: M_{(ab)} = 0 \}$ denote, respectively, the $N:=\frac{1}{2}
n(n-1)$-dimensional group of $g_{ab}$-orthogonal transformations and
the corresponding Lie algebra of anti-self dual infinitesimal
transformations.  
Let $\mM_\alpha=(\M^a_{b\alpha})$ be a basis of $\fg$, and let
$\C^\al_{\be\ga}$ denote the corresponding structure constants.
Henceforth, we use bivector indices $\al,\be,\ga=1,\ldots, N$ as
gauge indices. Let $\bG^\al$ denote the $\fg$-valued portion of the
connection 1-forms; to be more precise, $\bom_{[ab]} =
M_{ab\al}\bG^\al $.

Particular instances of a $\CH_0$ geometry are determined by an array
of constants $\tR^\al_{bc}=-\tR^\al_{cb}$.
The corresponding  $\CH_0$ field equations are as follows:
\begin{eqnarray}
  \label{eq:ch01}
  \bom_{(ab)} = 0,\\
  \label{eq:ch02}
  d\bom_{(ab)} = 0,\\
  \label{eq:ch03}
  d \bG^\al + \frac{1}{2} \C^\al_{\be\ga}\, \bG^\be \wedge \bG^\ga -
  \frac{1}{2} 
  \tR^\al_{bc}\, 
  \bom^b\wedge\bom^c=0,\\
  \label{eq:ch04}
  (\mM_\be\cdot \tilde{R})^\al_{\ bc}\,  \bG^\be\wedge\bom^b\wedge
  \bom^c  = 0, 
\end{eqnarray}
where
\begin{equation}
  (\mM_\be\cdot \tilde{R})^\al_{\ bc} =  \C^\al_{\be\ga} \tR^\ga_{bc} -  \M^a_{b\be}
  \tR^\al_{ac}- 
  \M^a_{c\be} \tR^\al_{ba} 
\end{equation}
is the standard action of $\fg$ on the vector space of
$\fg$-valued curvature-type tensors.  
Equations \eqref{eq:ch01} constrain the connection to be
$\fg$-valued; this is equivalent to the compatibility of
$\Gu^i_{jk}$ and $g_{ij}$. 
Equations \eqref{eq:ch02} and \eqref{eq:ch04} are the compatibility
conditions for \eqref{eq:ch01} and \eqref{eq:ch03}, respectively; one
takes the exterior derivative of the latter to obtain the former.  In
particular, equations \eqref{eq:ch02} constrain the curvature 2-form,
\[ \bOmu^a_b:= d\bomu^a_b + \bomu^a_c \wedge \bomu^c_b, \] to be
$\fg$-valued.  Equations \eqref{eq:ch04} express the differential
Bianchi identities. Equations \eqref{eq:ch03} (these are equivalent to
the NP equations in 4D relativity) together with \eqref{eq:ch02}
express the condition for curvature homogeneity, namely
\[ \bOm^a_{\ b} = \frac{1}{2}\M^a_{b\al} \tR^\al_{cd}\, \bom^c \wedge \bom^d.\] Hence, as a consequence of
\eqref{eq:taut2}, it is necessary to
assume that the $\tR^\al_{bc}$ satisfy the algebraic Bianchi identity,
namely
\begin{equation}
  \label{eq:algbianchi}
  \tR^\al_{[bc}\M^a_{d]\al}=0.
\end{equation}

\subsection{The $\CH_0$ class}
Set $G_{-1}:=G$ and $\fg_{-1}:=\fg$, and let $G_0\subseteq G_{-1}$ and
$\fg_0\subseteq \fg_{-1}$ be, respectively, the subgroup and the Lie
subalgebra of frame transformations that leaves invariant the constant
array $\tR^\al_{bc}$.  As a system, equations
\eqref{eq:ch01}-\eqref{eq:ch04} are invariant with respect to $G_0$.
Thus,  $G_0$ serves as the reduced gauge group of the
$\CH_0$ field equations.  Let us set $N_i=\dim\fg_i$ and arrange the
basis of $\fg$ so that $\mM_\la,\; \la=1,\ldots, N_{-1}-N_0$ is a
basis of $\fg_{-1}/\fg_0$ and so that $\mM_\xi,\;
\xi=N_{-1}-N_0+1,\ldots, N_{-1}$ is a basis of the reduced subalgebra
$\fg_0$.  It is important to note that if $\fg_0=\fg_{-1}$, then the
Bianchi identities \eqref{eq:ch04} reduce to a trivial constraint.
Consequently, equations \eqref{eq:taut1} and \eqref{eq:ch03} function
as the structure equations of a homogeneous space --- in this case a
space of constant curvature.  Thus, to obtain a proper $\CH_0$
geometry it is necessary to assume that the inclusion $\fg_0\subset
\fg_{-1}$ is proper.

The components of the connection forms $\bG^\al$ do not obey a
meaningful transformation law with respect to the full group $G$.
However, once we restrict our group to $G_0$, we distinguish between
two different kinds of connection forms : $\bG^\xi$, whose components
are not $G_0$-equivariant, and
\begin{equation}
  \label{eq:ch05}
  \bG^\la = \Gu^\la_a\, \bom^a,
\end{equation}
whose $G_0$-equivariant components $\Gu^\la_a$ we treat as new scalar
variables.  By \eqref{eq:omij}, these new variables represent certain
linear combinations of the first order jets
\[\u^i_{a;b}=\partial_j \u^i_a \u^j_b-\u^i_{ab}.\] 
The $\Gu^\la_a$ obey the following field equations:
\begin{eqnarray}
  \label{eq:ch06}
  d\Gu^\la_a \wedge \bom^a + (\mM_\xi \cdot \Gamma)^\la_{\
    a}\ \bG^\xi \wedge \bom^a+\frac{1}{2}
  \Up^\la_{bc}\ \bom^b\wedge \bom^c = 0, 
\end{eqnarray}
where
\begin{eqnarray}
  \label{eq:Updef}
  \Up^\la_{bc} =  -2\, \Gu^\la_a\, \Gu^\mu_{[b}\ \M^a_{c]\mu} +
  \Gu^\mu_b\, \Gu^\nu_c\,\C^\la_{\mu\nu} - \tR^\la_{bc}.
\end{eqnarray}
Here
\begin{eqnarray}
  (\mM_\xi\cdot \Gamma)^\la_{\ a} = \C^\la_{\xi\mu} \Gu^\mu_b -
  \M^b_{a\xi} \Gu^\la_b
\end{eqnarray}
denotes the natural action of $\fg_0$ on $\cL(\Rset^n,\fg_{-1}/\fg_0)$,
the space of linear maps from $\Rset^n$ to $\fg_{-1}/\fg_0$.  Eqns.
\eqref{eq:ch06} follow from \eqref{eq:ch03} and from the exterior
derivative of \eqref{eq:ch05}.

Taken together, \eqref{eq:ch01}-\eqref{eq:ch04} generate an exterior
differential system.  The total space is $\cF^2 M$, the bundle of \sec
order frames over $M$ \cite{Kobayashi72}.  The variables $u^i, \u^i_a,
\u^i_{bc}$ serve as local coordinates on $\cF^2M$.  We impose the
independence condition
\begin{equation}
  \label{eq:indcond1}
  \bom^1\wedge\cdots \wedge\bom^n\neq 0,
\end{equation}
which means that $u^i$ are to be regarded as independent variables,
and $\u^i_a, \u^i_{bc}$ as the field variables.  An $n$-dimensional
integral manifold with independence condition \eqref{eq:indcond1}
describes a $g_{ab}$-orthogonal moving frame with constant
$\R^\al_{bc}$.  From a theoretical point of view,
however, it is more 
advantageous to consider $(n+N_0)$-dimensional integral manifolds with
independence condition
\begin{equation}
  \label{eq:indcond2}
  (\bigwedge_a \om^a) \wedge (\bigwedge_\xi\bG^\xi)\neq 0.
\end{equation}
The latter type of integral manifold describes a $\CH_0$ bundle of
$g_{ab}$-orthogonal frames with reduced structure group $G_0$.

For general choices of the $\tR^\al_{bc}$ the EDS is not
involutive.  We must (partially) prolong the
system by introducing the extra field variables $\Gu^\la_a$; the
prolonged total space is $\cF^2M \times \cL(\Rset^n,\fg_{-1}/\fg_0)$.
The prolonged EDS is generated by equations
\eqref{eq:ch01}-\eqref{eq:ch04}, \eqref{eq:ch05}, \eqref{eq:ch06}.
Using \eqref{eq:ch05}, the Bianchi identities take the form of linear
constraints
\begin{equation}
  \label{eq:pbianchi}
  (\mM_\la \cdot \tilde{R})^\al_{\ [bc} \Gu^\la_{a]} =0.
\end{equation}
Some linear combinations of \eqref{eq:ch06} reduce to algebraically quadratic
constraints on the $\Gu^\la_a$.  In EDS theory, such algebraic
constraints are called torsion \cite{IvLa03}.  In order for the
prolonged equations to have solutions, it is necessary to constrain
the prolongation variables so that all torsion vanishes.



\subsection{Higher order curvature homogeneity}
On a manifold of class $\CH_1$, the curvature tensor $\R^\al_{bc}$ and
and its covariant derivative $\nabla_a \R^\al_{bc}$ are constant
relative to some choice of moving frame.  Consequently, there is a
chain of 3 Lie algebras, call them $\fg_1\subseteq \fg_0\subseteq
\fg_{-1}$, such that $\R^\al_{bc}=\tR^\al_{bc}$ and $\nabla_a
\R^\al_{bc}=\tR^\al_{abc}$, where the latter are
respectively, a $\fg_0$-invariant and a $\fg_1$-invariant
array of constants.  We also have 3 groups of generators: $\mM_\rho$
a basis of $\fg_{-1}/\fg_0$, $\mM_\la$ a basis of $\fg_0/\fg_1$, and
$\mM_\xi$ a basis of $\fg_1$.  In other words, a choice of
$\tR^\al_{bc}$ reduces the gauge/structure
group from $\fg_{-1}$ to $\fg_0$; then, $\tR^\al_{abc}$
further reduces the group to $\fg_1$.

Since the $\tR^\al_{bc}$ are constant, we have
\begin{equation}
  \label{eq:DR1}
  \tR^\al_{abc} = (\mM_\rho \cdot \tilde{R})^\al_{\ bc}\,\Gu^\rho_a.
\end{equation}
Since the $\mM_\xi, \mM_\la$ span the kernel of the action of $\fg_{-1}$
on $\tR^\al_{bc}$, by definition, one can fully back-solve the linear
system \eqref{eq:DR1} to obtain $\Gu^\rho_a$ as a bilinear combination
of $\tR^\al_{bc}$ and $\tR^\al_{abc}$.  In describing a $\CH_1$
geometry, it therefore suffices to specify a $\fg_1$-invariant array
of constants, $\Gu^\rho_a=\tGu^\rho_a$, rather than
an array of constants $\tR^\al_{abc}$.

The EDS for
$\CH_1$ geometry is given by equations
\eqref{eq:ch01}-\eqref{eq:ch04}, and by
\begin{eqnarray}
  \label{eq:ch07}
  \bG^\rho - \tGu^\rho_a\, \bom^a =0\\
  \label{eq:ch08}
  (\mM_\la \cdot \tGamma)^\rho_{\
    a}\ \bG^\la \wedge \bom^a+\frac{1}{2}
  \tUp^\rho_{bc}\ \bom^b\wedge \bom^c = 0,
\end{eqnarray}
where $\tUp^\rho_{bc}$ are constants that are given, mutatis mutandi,
by \eqref{eq:Updef}.  
The Bianchi equations \eqref{eq:pbianchi}  act as linear constraints
on the $\tGu^\rho_a$.  Certain linear combinations of \eqref{eq:ch08} may
act as quadratic constraints.
Even if these constraints are satisfied, the resulting EDS is, in
general, not involutive; we have to prolong.  As above, we introduce
additional scalar variables $\Gu^\la_a$ and additional field equations
\eqref{eq:ch05} and \eqref{eq:ch06}.  As a consequence of
\eqref{eq:ch05} and \eqref{eq:ch08},  we have the linear
constraints
\begin{equation}
  \label{eq:ch08a}
   (\mM_\la \cdot \tGamma)^\rho_{\
    [b}\ \Gu^\la_{c]} =\frac{1}{2}
  \tUp^\rho_{bc}\ .
\end{equation}
As well, some linear combinations of the field equations
\eqref{eq:ch06} reduce to algebraically quadratic constraints on the
$\Gu^\la_a$.  Thus, it is necessary to constrain the $\tGu^\rho_a$ and
$\Gu^\la_a$ so that all torsion implied by \eqref{eq:pbianchi},
\eqref{eq:ch06}, \eqref{eq:ch08}, \eqref{eq:ch08a} vanishes.  We
define a $\CH_1$ configuration to be a triple
$(\tR^\al_{bc},\tGu^\rho_a, V_1)$, where the $\tR^\al_{bc}$ is an
array of $\fg_0$-invariant constants that satisfies
\eqref{eq:algbianchi}, where $\tGu^\rho_a$ is an array of
$\fg_1$-invariant constants, and where $V_1\subseteq
\cL(\Rset^n,\fg_0/\fg_1)$ is an algebraic variety such that the EDS
generated by \eqref{eq:ch01}-\eqref{eq:ch04}, \eqref{eq:ch05},
\eqref{eq:ch06}, \eqref{eq:ch07}, \eqref{eq:ch08} on $\cF^2M \times
V_1$ is torsion-free.

An important subcase arises when $\fg_1=\fg_0$ \cite{FeRe06}.  In this
case, the constants $\tGu^\rho_a$ must be $\fg_0$-invariant, and must
satisfy the linear and quadratic constraints
\[ (\mM_\rho \cdot R)^\al_{\ [bc} \tGu^\rho_{a]}=0,\quad
\tUp^\rho_{bc}=0.\] Since $\fg_0/\fg_1$ is trivial, no additional
constraints can or need be imposed.  The unprolonged EDS on $\cF^2M$,
generated by \eqref{eq:ch01}-\eqref{eq:ch04}, \eqref{eq:ch07} is
involutive.  The $(n+N_0)$-dimensional integral manifolds of this EDS
are homogeneous spaces with $\fg_0$ isotropy and structure equations
\begin{eqnarray}
  \label{eq:ch1seq1}
  d\bom^a+ \M^a_{c\rho} \tGu^\rho_b\, \bom^b\wedge\bom^c +
  \M^a_{b\xi} \bG^\xi\wedge \bom^b = 0\\
  \label{eq:ch1seq2}
  d \bG^\xi + \frac{1}{2} \C^\xi_{\upsilon\zeta} \bG^\upsilon \wedge
  \bG^\zeta - \frac{1}{2} 
  \tR^\xi_{bc}\, 
  \bom^b\wedge\bom^c=0.
\end{eqnarray}
We say that a $\CH_1$ configuration is proper if the inclusion
$\fg_1\subset\fg_0$ is proper.
In seeking proper $\CH_1$ geometries one must demand that the $\CH_1$
configuration be proper.

The general treatment of $\CH_s,\; s\geq 1$ is similar.  Now one has
to consider a chain of Lie algebras
$\fg_{s}\subseteq\cdots\subseteq\fg_0\subseteq\fg_{-1}$, with
$\mM_{\rho_r},\; r=1,\ldots, s$ a basis of $\fg_{r-2}/\fg_{r-1}$, with
$\mM_\la$ a basis for $\fg_{s-1}/\fg_{s}$, and $\mM_\xi$ a basis for
$\fg_{s}$. The linear coordinates, $\Gu^\la_a$, on
$\cL(\Rset^n,\fg_{s-1}/\fg_{s})$, are prolongation variables.  A
$\CH_s$ configuration is an $(s+2)$-tuple
$(\tR^\al_{bc},\tGu^{\rho_r}_a, V_s)$, where $\tR^\al_{bc}$ is an
array of $\fg_0$-invariant constants, where each $\tGu^{\rho_r}_a$ is
an array of $\fg_{r}$-invariant constants, and where
$V_s\subseteq\cL(\Rset^n,\fg_{s-1}/\fg_{s})$ is an algebraic variety
such that the EDS generated by \eqref{eq:ch01}-\eqref{eq:ch04},
\eqref{eq:ch05}, \eqref{eq:ch06}, \eqref{eq:ch07}, \eqref{eq:ch08} on
$\cF^2M\times V_s$ is torsion-free.  The definition of involutive
configuration is as above.  An $n$-dimensional integral manifold with
independence condition \eqref{eq:indcond1} describes a
$g_{ab}$-orthogonal frame with constant $\nabla_{c_1\ldots
  c_r}\R^a_{b_1 b_2 b_3},\; r=0,1,\ldots, s$.  An
$(n+N_{s})$-dimensional integral manifold with independence condition
\eqref{eq:indcond2} describes a bundle of $g_{ab}$-orthogonal frames
with reduced structure group $G_{s}$.

\section{Proper $\CH_2$ geometries}

Henceforth, we assume that $M$ is four-dimensional and that $g_{ab}$
is the Lorentzian inner product.  We will express our calculations
using the NP formalism, which is based on complex null
tetrads $\{ \ve_a\} = (\vm, \bvm, \vn, \vell)$, i.e., the metric is
given by
\[ g_{ij}\, du^i du^j = \bom^1\bom^2 - \bom^3\bom^4,\quad \bom^2 =
\overline{\bom^1},\; \bom^3 = \overline{\bom^3},\;
\bom^4=\overline{\bom^4}.\] The connection components are labeled by
NP spin coefficients:
\begin{eqnarray*}
  &-\bom_{14}= \sigma\bom^1+\rho\bom^2+\tau\bom^3 + \kappa\bom^4;\quad
  \bom_{23} = \mu\bom^1+\la \bom^2+\nu\bom^3 +\pi\bom^4;\\
  &-\frac{1}{2}(\bom_{12}+\bom_{34}) = \be \bom^1 +
  \alpha\bom^2+\ga\bom^3+\ep\bom^4.
\end{eqnarray*}
The curvature scalars $\R^\al_{bc}$ are given by $\Lambda, \Phi_{AB},
\Psi_C$, in the usual manner \cite{ES}.

A proper $\CH_2$ configuration requires a chain of proper inclusions
$\fg_2\subset\fg_1\subset\fg_0\subset\fg_{-1}$ so that $\fg_0$ is the
symmetry algebra of $\tR^\al_{bc}$, and $\fg_1,\fg_2$ are,
respectively, the symmetry algebras of $\tGu^{\rho_1}_a,
\tGu^{\rho_2}_a$.  If the Petrov type is I, II, or III, then one can
fully fix the frame by setting $\Psi_0=\Psi_4=0$ and then normalizing
$\Psi_1$ or $\Psi_3$ to $1$.  In other words, $N_1=N_0=0$, and hence,
by Singer's criterion,  every $\CH_1$ manifold of Petrov type I, II,
or III is  a homogeneous space.   Proper $\CH_1$ Lorentzian manifolds
must, necessarily, be of Petrov type D, N, or O.  These will be
studied in a future work.

A proper $\CH_2$ configuration, if one exists, can potentially arise
in 5 ways.  Only in  case 5.2, do we obtain a proper $\CH_2$
configuration.  A proper $\CH_3$ configuration, if one exists, is
limited to the first 3 cases.  However, since in each case we rule out
the existence of a proper $\CH_2$ configuration, a proper $\CH_3$
configuration does not exist either.

\noindent
1) $\fg_0=\so(3)$.  The curvature is that
of a conformally flat perfect fluid.
\begin{eqnarray*}
  \Psi_A= \Phi_{01} =
  \Phi_{02}=\Phi_{12} = 0.\quad \Phi_{00}=\Phi_{22}=2 \Phi_{11}\\
  \tGu^{\rho_1}_a \in \mathrm{span} \{ \kappa-\bar{\pi},\ 
  \sigma-\bar{\la},\ \rho-\bar{\mu},\ \tau-\bar{\nu},\ \al+\bar{\be},\
  \ep_1:=\Re(\epsilon),\ \ga_1:=\Re(\ga)\ \}.
\end{eqnarray*}
Bianchi identities imply that all $\tGu^{\rho_1}_a=0$. Necessarily,
$\fg_1=\fg_0$. A proper configuration does not exist.

\noindent
2) $\fg_0=\so(1,2)$. The curvature constants are
\begin{eqnarray*}
  \Psi_A= \Phi_{01} =
  \Phi_{02}=\Phi_{12} = 0.\quad \Phi_{00}=\Phi_{22}=-2 \Phi_{11}\\
  \tGu^{\rho_1}_a \in \mathrm{span} \{ \kappa+\bar{\pi},\
  \sigma+\bar{\la},\ \rho+\bar{\mu},\ \tau+\bar{\nu},\ \al+\bar{\be},\
  \ep_1,\ \ga_1\ \}.
\end{eqnarray*}
Again, by Bianchi $\fg_1=\fg_0$.  There are no proper configurations.

\noindent 3)
$\fg_0$ is three-dimensional, generated by
spins and null rotations.  The curvature is that of an aligned null
radiation field on a conformally flat background:
\begin{eqnarray*}
  \Psi_A=\Phi_{00}=\Phi_{01}=\Phi_{11}=\Phi_{02}=\Phi_{12} = 0.\\
  \tGu^{\rho_1}_a \in \mathrm{span} 
  \{\kappa,\sigma,\rho,\tau,\al+\bar{\be},\ga_1,\ep_1 \}. 
\end{eqnarray*}
The Bianchi identities imply $\kappa=\sigma=\rho=0$,
$\tau=2\bar{\al}+2\be$.  If $\tau=0$, then $\fg_1=\fg_0$.  Performing
a spin, as necessary, suppose, WLOG $\tau=\bar{\tau}\neq 0$. Hence,
$\fg_1$ is generated by imaginary null rotations; $\tGu^{\rho_2}_a \in
\mathrm{span}
\{\al-\bar{\be},\ga_2:=\Im(\ga),\ep_2:=\Im(\ep),\la+\bar{\mu},\nu_1,\pi_1
\}$.  The NP  
equations for a $\CH_1$ configuration give $\beta=-\frac{1}{4}\tau$,
$\alpha=\frac{3}{4} \tau$, $\pi_1=-\tau$, $\ep_2=0$, $\Lambda= -24 \tau^2$,
$\lambda+\bar{\mu} = -\frac{2}{3} \ga$.  These constraints describe a
proper $\CH_1$ configuration.  However, the additional NP equations
for a proper $\CH_2$ configuration, imply $\Phi_{22} = \frac{8}{9}
\ga_1^2 - 2 \nu_1 \tau$, $\ga_2=0$.  The last condition implies that
$\fg_2=\fg_1$.  Therefore, a proper $\CH_2$ configuration does not
exist.

\noindent 4) $\fg_0$ is generated by spins and boosts,
$N_0=2$.  Petrov type is D/O.
\begin{eqnarray*}
\Psi_0=\Psi_1=\Psi_3=\Psi_4=\Phi_{00} = \Phi_{01} =
\Phi_{02}=\Phi_{12}= \Phi_{22} = 0.\\
\tGu^{\rho_1}_a \in \mathrm{span} \{\kappa, \sigma, \rho, \tau,
\lambda, \mu, \nu, \pi\}.
\end{eqnarray*}
We require $N_1=1$. There are 2 subcases.  Neither admits a proper
configuration.

4.1) $\fg_1=$ spins, $\kappa=\sigma=\lambda=\nu=\tau=\pi=0$, $(\rho,\mu)\neq (0,0)$.  Bianchi
identities imply $\Psi_2 = -\frac{2}{3} \Phi_{11}$, $\mu+\bar{\mu}=0$,
$ \rho+\bar{\rho}=0$.  The NP equations, then imply $\rho^2=0$, and
$\mu^2=0$, a contradiction.

4.2)
$\fg_1=$ boosts,
$\kappa=\sigma=\lambda=\nu=\rho=\mu=0$, 
$(\tau,\pi)\neq (0,0)$.  By Bianchi,
$\Psi_2= \frac{2}{3}\Phi_{11}$, $\tau=\bar{\pi}$. NP equations
imply $\pi^2 = 0$, a contradiction.

\noindent
5) 
$\fg_0$ is 2-dimensional, generated by
null rotations.  The curvature is null radiation/vacuum with an
aligned type N or conformally flat background:
\begin{eqnarray*}
 \Psi_0=\Psi_1=\Psi_2=\Psi_3=\Phi_{00} = \Phi_{01} = \Phi_{11} =
 \Phi_{02} = \Phi_{12} = 0; \\
 \tGu^{\rho_1}_a \in \mathrm{span} \{\kappa, \sigma, \rho, \tau,
 \alpha, \beta, \gamma, \epsilon\}.
\end{eqnarray*}
To obtain a proper $\CH_2$ configuration, we require $N_1=1$.  WLOG,
$\fg_1$ consists of imaginary null rotations; $\tGu^{\rho_2}_a \in
\mathrm{span} \{\la+\bar{\mu},\ \nu_1,\ \pi_1\}$.  This requires
$\kappa=\sigma=\rho=0$, $\alpha=\beta+\tau\neq 0$.  We use
a real null rotation to set $\ga_1=0$. There are two subcases: the
singular case $\ga_2=0$, and the generic case $\ga_2\neq 0$.

5.1)
Suppose $\ga_2=0$. The $\CH_1$ NP
equations give $\tau_2=\be_2=0$, $\Lambda=-24\tau^2$, $\bar{\Psi}_4 =
\Psi_4$, $\pi_1=-\tau_1$, $\ga_2=0$, $\la=-\bar{\mu}$.  However, the
last 3 equations imply $\fg_2=\fg_1$.  Therefore, no proper $\CH_2$
configurations are possible.

5.2) Suppose $\ga_2\neq 0$. The $\CH_1$ NP
equations and Bianchi identities give $\tau_2=0$, $\be=\tau_1/2$,
$\Lambda=-24\tau_1^2$, $\Psi_4=3\Phi_{22}$, $\pi_1=-\tau_1$,
$\la=-\bar{\mu}-\frac{2i}{3} \ga_2$.  We ask that $N_2=0$;
$\Gu^\la_a\in \mathrm{span}(\la-\bar{\mu}, \nu_2,\pi_2\ \}$.
This requires $\ga_2\neq 0$.  WLOG, an imaginary null rotation
normalizes $\nu_2=0$, and a boost normalizes $\ga_2=\frac{3}{2}$.  The
$\CH_2$ NP equations then imply $\pi_2=0$, $\mu=\frac{2i}{5}(4+
\Phi_{22})$.

The EDS for the resulting proper $\CH_2$ configuration is generated by
a Pfaffian system (1-forms and their exterior derivatives),  
\begin{eqnarray*}
  \fl \bom_{(ab)} = 0,\quad \bom_{[14]}= -\tau_1 \bom^1,\;
  \bom_{[34]} = 
  -2\tau_1(\bom^1+\bom^2),\;
  \bom_{[12]}=\tau_1(\bom^1-\bom^2)-3 i \bom^3,\\
  \fl \bom_{[13]}=i(\bom^1-\bom^2)-\tau_1\bom^4-\frac{i}{5}(3+2\Phi_{22})(\bom^1
  +\bom^2)-i\nu_2\,\bom^3 
\end{eqnarray*}
and an additional 2-form,
\[ \fl d\nu_2\wedge \bom^3 = \frac{3i}{5} (4+\Phi_{22})
(\bom^1-\bom^2)\wedge \bom^3 - 3\nu_2 \tau_1
(\bom^1+\bom^2)\wedge\bom^3+3\tau_1 \bom^3\wedge\bom^4.\] Here
$\tau_1\neq 0, \Phi_{22}\neq 0$ are real constants.  The total space
is $\cF^2 M\times \Rset$, with $\nu_2$ the coordinate of the extra
factor.  Cartan's test yields $s_0 = 16$, $s_1 = 17$, $s_2=12$, $s_3 =
8$, $s_4=4$, which proves that the EDS is involutive.  Subtracting
(16,16,12,8,4), the parameter counts corresponding to a choice of
local coordinates, we infer that, up to a local diffeomorphism, the
general solution depends on 1 function of 1 variable.  Indeed, an
exact solution can be given:
\begin{eqnarray}
  \label{eq:ch2om1}
  \fl \bom^1 = \frac{1}{2 C_1} \left( (1+3 iu)  dr +3i
    du-e^r \left( 3u + i 
      \left( C_1^2 C_2+
        \frac{9}{2}u^2-\frac{5}{2}\right)\right)dx\right)\\  
  \fl  \bom^3 = e^r dx\\
  \label{eq:ch2om4}
  \fl \bom^4 = dv+3(C_2u + v) dr + 
  \left(f(x) e^{-3r} +e^r \left(\frac{C_1^2 C_2^2}{4} - 9
      uv-\frac{5}{8}C_2(1+18u^2)\right)\right)dx
\end{eqnarray}
where $x,r,u,v$ are coordinates; $f(x)$ is the functional parameter;
$C_1, C_2$ are real constants such that $\tau_1=C_1$, $\Phi_{22} = -4
+ \frac{5}{2} C_1^2 C_2$, $\nu_2 = -\frac{3}{2}C_1(3 C_2 u +2v)$.

Finally, we note that, since $N_2=0$, it is impossible to refine the
above to a proper $\CH_3$ configuration.

\ack
The research of RM is supported in part by NSERC grant RGPIN-228057-2004.

\end{document}